\documentclass[12pt,preprint]{aastex}

\shorttitle{Radiative transfer in dust grain mixtures}
\shortauthors{S. Wolf}

\begin{document}

%% LaTeX will automatically break titles if they run longer than
%% one line. However, you may use \\ to force a line break if
%% you desire.

\title{Efficient radiative transfer in dust grain mixtures}

%% Use \author, \affil, and the \and command to format
%% author and affiliation information.
%% Note that \email has replaced the old \authoremail command
%% from AASTeX v4.0. You can use \email to mark an email address
%% anywhere in the paper, not just in the front matter.
%% As in the title, you can use \\ to force line breaks.

\author{S. Wolf}
\affil{California Institute of Technology, 1200 E California Blvd,\\
Mail code 220-6, Pasadena, CA 91125}

%% Notice that each of these authors has alternate affiliations, which
%% are identified by the \altaffilmark after each name.  Specify alternate
%% affiliation information with \altaffiltext, with one command per each
%% affiliation.

%\altaffiltext{1}{Patron, Alonso's Bar and Grill}

%% Mark off your abstract in the ``abstract'' environment. In the manuscript
%% style, abstract will output a Received/Accepted line after the
%% title and affiliation information. No date will appear since the author
%% does not have this information. The dates will be filled in by the
%% editorial office after submission.

\begin{abstract}
The influence of a dust grain mixture consisting 
of spherical dust grains with different radii and/or chemical composition 
on the resulting temperature structure and spectral energy distribution 
of a circumstellar shell is investigated. The comparison with the results based on
an approximation of dust grain parameters representing the mean optical properties 
of the corresponding dust grain mixture reveal that
(1) the temperature dispersion of a real dust grain mixture decreases substantially
with increasing optical depth, converging towards the temperature distribution
resulting from the approximation of mean dust grain parameters, and
(2) the resulting spectral energy distributions do not differ by more than 10\%
if $\ge 2^5$ grain sizes are considered
which justifies the mean parameter approximation and the many results obtained
under its assumption so far.
Nevertheless, the dust grain temperature dispersion at the inner boundary
of a dust shell may amount to $\gg$100\,K and has therefore to be considered
in the correct simulation of, e.g., chemical networks.

In order to study the additional influence of geometrical effects, a two-dimensional
configuration -- the HH\,30 circumstellar disk -- was considered, using model 
parameters from Cotera et al.~(2001) and Wood et al.~(2002). A drastic inversion
of the large to small grain temperature distribution was found within the inner
$\sim 1{\rm AU}$ of the disk.
\end{abstract}

%% Keywords should appear after the \end{abstract} command. The uncommented
%% example has been keyed in ApJ style. See the instructions to authors
%% for the journal to which you are submitting your paper to determine
%% what keyword punctuation is appropriate.


%% From the front matter, we move on to the body of the paper.
%% In the first two sections, notice the use of the natbib \citep
%% and \citet commands to identify citations.  The citations are
%% tied to the reference list via symbolic KEYs. The KEY corresponds
%% to the KEY in the \bibitem in the reference list below. We have
%% chosen the first three characters of the first author's name plus
%% the last two numeral of the year of publication as our KEY for
%% each reference.

%---------------------------------------------------------------------------------------------------
\section{Introduction}\label{intro}

The simulation of spectral energy distributions (SEDs), images, and polarization maps of
young stellar objects has become a profound basis for the analysis and interpretation
of observing results. Many techniques and approximations for the solution
of the radiative transfer (RT) problem in different model geometries (1D--3D),
considering more and more special physical processes, such as
the stochastic and photo-electric heating of small grains 
(see, e.g, Draine \& Li~2001;
Bakes \& Tielens~1994, Siebenmorgen, Kr\"ugel, \& Mathis~1992),
scattering by spheroidal grains 
(see, e.g., Wolf, Voshchinnikov, \& Henning~2002; Gledhill \& McCall~2000),
or the  coupling of line and continuum RT (see, e.g., Rybicki \& Hummer~1992), have been developed.
The simulation of the temperature structure in simple-structured circumstellar shells or disks 
(see, e.g., Malbet, Lachaume, \& Monin~2001; Chiang et al.~2001), 
the estimation of the properties (luminosity, temperature, mass)
of heavily embedded stars (see, e.g., Kraemer et al.~2001), 
or the determination of the inclination of a
circumstellar disk 
(see, e.g., Chiang \& Goldreich~1999;
Men'shchikov, Henning, \& Fischer~1999;
Wood et al.~1998) 
represented modest, first attempts of the application
of the existing sophisticated numerical techniques.
More recent efforts are directed
to derive the dust grain size distribution in a circumstellar disk 
from its SED (Wood et al.~2002; D'Alessio, Calvet \& Hartman~2001). 
Thus, it is clear that beside strong observational
constrains, the model parameters and considered physical processes have to be questioned
in depth in order to derive such detailed conclusions.
However, looking behind the scenes, many of the RT models
are based on simplifying assumptions of very basic processes such as isotropic instead
of anisotropic scattering, mean dust parameters representing dust grain ensembles
(different radii and chemical compositions), or the flux-limited diffusion approximation
- approximations which are well-suited for handling the energy transfer in hydrodynamic
simulations or a rough data analysis but which may not necessarily guarantee the desired
accuracy for a detailed SED and image/polarization data analysis.

In the advent of (space) observatories such as SIRTF\footnote{Space Infrared Telescope Facility},
which will be able to obtain SEDs of evolved debris disks around young stars, there exists
a strong need for adequate numerical RT techniques in order to allow to trace dust 
grain growth (Meyer et al.~2001) and the influence of other physical effects 
and processes such as the Poynting-Robertson effect (see, e.g., Srikanth~1999) 
and dust settling (Dubrulle, Morfill, \& Sterzik~1995, Miyake \& Nakagawa~1995) 
on the dust grain size distribution in these disks.
Therefore, we clearly have to understand which influence the different approximations
(as long as they are required) in the RT simulations may have on the resulting observables.
Based on two different grain size distributions consisting of astronomical silicate,
the differences in the resulting dust grain temperature distributions and the
resulting SEDs between simulations of the RT in (A) a ``real'' dust grain mixture and 
(B) under the assumption (approximation) of mean dust grain parameters will be discussed.
This investigation is therefore mainly focused on two questions:
(1) Of what order of magnitude are the differences (?) and
(2) How many grain sizes have to be considered to represent the properties of a real
grain size distribution?
The spatial temperature distribution of the considered dust configurations is
calculated on the basis of local thermal equilibrium. Stochastic heating processes
which are expected in case of very small grains consisting of tens to hundreds of atoms
(see, e.g., Draine \& Li~2000 and references therein)
are not subject of this investigation.

In Sect.~\ref{rtmodel}, the RT and the dust grain model are briefly introduced.
In Sect.~\ref{mean}, the definition of the mean dust grain parameters is given
and the expected deviations of the RT results -- once based on the mean dust parameters and
once on a real grain size distribution -- are outlined.
In Sect.~\ref{rtem}, the RT in a spherical shell with variable
optical depth and density distribution (see Sect.~\ref{spsh}) is considered,
while the temperature structure in a model of the HH\,30 circumstellar disk
is investigated in Sect.~\ref{hh30}.
The SED resulting from a models with a dust grain mixture is compared to the mean
particle approximation in Sect.~\ref{sed}.

%---------------------------------------------------------------------------------------------------
\section{Radiative transfer and Dust model}\label{rtmodel}

The RT simulations presented in this article
have been performed with the three-dimensional continuum RT code MC3D
which has been described by Wolf \& Henning~2000 (see also Wolf, Henning, \& Stecklum~1999; Wolf~2002).
It is based on the Monte-Carlo method and solves the RT problem self-consistently.
Instead of the iterative procedure of estimating the dust temperature distribution
(as described in Wolf et al.~1999), the concept presented by Bjorkman \& Wood~(2001),
which is based on the immediate correction of the dust grain temperature after absorption
of a photon package, has been used. Furthermore, the method described by Lucy~(1999),
which takes into account the absorption not only at the points of interaction of the photons
with the dust grains but also in-between, has been applied. While the first method
allows to simulate the RT for optical depths\footnote{Throughout the whole article the quantity
$\tau$ refers to the optical depth based on the extinction cross section 
of the dust grains at a wavelength of 550\,nm.} $\tau \gg 10^3$, the second was used
in order to increase the efficiency of the simulation for low optical depths.
In addition to the test of MC3D described by Wolf et al.~(1999) considering a single
grain size, it has been successfully tested for the case of dust grain mixtures against the 
one-dimensional code of Chini et al.~(1986).

The RT is simulated at separate wavelengths within the wavelength
range [$\lambda_{\rm min}$, $\lambda_{\rm max}$]. For this reason, the radiation energy
of the emitting source(s) is partitioned into so-called
weighted photons (Fischer, Henning, \& Yorke~1999) each of which is characterized
by its wavelength $\lambda$ and Stokes vector $\hat{I}=(I,Q,U,V)^{\rm T}$.
In order to consider dust grain ensembles instead of a single grain species,
the RT concept as described by Wolf et al.~(1999), Lucy~(1999), and Bjorkman \& Wood~(2001)
has to be extended in the following manner:
\begin{enumerate}
\item The mean free path length $l$ between the point of emission and the point of the first
photon-dust interaction (and - subsequently - between two points of interaction) is given by
\begin{equation}\label{tau-ext}
  \tau_{\rm ext,l} = -\ln(1-Z),
\end{equation}
\begin{equation}\label{sums}
  \tilde{\tau}_{\rm ext,l}
  = 
  \sum^{n_{\rm P}}_{i=1}
  \left[
  \sum^{n_{\rm D}}_{j=1}
  \sum^{n_{\rm S}}_{k=1}
  \rho_{j,k}(\vec{r}_{\rm l_i}) \, C_{\rm ext_{j,k}}
  \right]
  \, \Delta l_i,
  \hspace*{1cm} l = \sum^{n_{\rm P}}_{i=1} \Delta l_i.
\end{equation}
Here, $\tau_{\rm ext,l}$ is the optical depth along the path of the length $l$, 
$\tilde{\tau}_{\rm ext,l}$ is the corresponding numerical accumulated optical depth
($\tau_{\rm ext,l} - \tilde{\tau}_{\rm ext,l} \rightarrow 0$),
$n_{\rm D}$ is the number of chemically different dust grain species,
$n_{\rm S}$ is the number of particle radii being considered, and
$\vec{r}_{\rm l_i}$ is the spatial coordinate corresponding to the $i$th
integration point along the path length~$l$.
The quantity $\rho(\vec{r})$ 
is the number density at the spatial coordinate $\vec{r}$, the quantity
$C_{\rm ext}$ is the extinction cross section, and
$n_{\rm P}$ is the number of integration points along the path. 
Furthermore, $Z$ is a random number uniformly distributed in the interval $[0,1]$. 

\item According to the concept of immediate reemission described by Bjorkman \& Wood~(2001),
at each point of interaction of a photon with a dust grain either scattering or absorption
occurs. The probability for a photon to undergo the one or the other interaction process
with a grain with the chemical composition \#$j$ and the particle radius \#$k$
is given by
\begin{equation}
  \Pi_{x_{j,k}}(\vec{r}) = 
  \frac{\rho_{j,k}(\vec{r}) \, C_{x_{j,k}}}
       {
	 \sum_{m=1}^{n_{\rm D}} 
	 \sum_{p=1}^{n_{\rm S}}
	 \left[\rho_{m,p}(\vec{r}) \, C_{ext_{m,p}}\right]
       },
\end{equation}
where '$x$' stands either for absorption or scattering.
In case of absorption, the immediate reemission occurs from the same dust grain (species).

\item In Lucy's concept (1999), which considers the absorption of the electromagnetic radiation
field not only at the end points of the photon path (points of interaction) but also in-between,
simply the absorption due to all dust grain species (instead of only one) has to be taken
into account:
\begin{equation}
  \tau_{\rm abs} = 
  \sum_{j=1}^{n_{\rm D}} 
  \sum_{k=1}^{n_{\rm S}} 
  \int_{\rm point_1}^{\rm point_2} 
  C_{\rm abs_{j,k}} \, \rho_{j,k}(\vec{r})\ dr.
\end{equation}

\end{enumerate}

The investigations presented in this paper are based on dust grain ensembles
with different grain size distributions but the same chemical composition.
Since {\em the formalism (and therefore at least the qualitative conclusions) are the same
for ensembles consisting of grains with different chemical composition or size distribution
or even both}, this restriction is justified. The optical properties of astronomical
silicate (Draine \& Lee~1984) for [A] a size distribution of small grains with radii
$a=(0.005-0.25)\,\mu$m and [B] a size distribution of larger grains with 
radii $a=(0.1-1)\,\mu$m have been used in Sect.~\ref{mean}, \ref{spsh}, and \ref{sed}.

The grains are assumed to be spherical with
a size distribution described by the widely applied power law $n(a) \propto a^{-3.5}$ 
(Mathis, Rumpl, \& Nordsieck~1977).
The optical properties such as the extinction and scattering cross sections as well as
the scattering distribution function have been derived on the basis of Mie scattering,
using the Mie scattering algorithm published by Bohren \& Huffman (1983). The correct
(anisotropic) scattering distribution function (for each dust grain size) has
been considered in the RT process.

%---------------------------------------------------------------------------------------------------
\section{Mean dust grain parameters}\label{mean}

In case of non-self-consistent RT in a dust grain mixture,
the numerical effort (finally, the computing time and RAM requirement) may be substantially
decreased by assuming weighted mean values of those parameters which describe
the interaction of the electromagnetic field with the dust grains:
the extinction, absorption, and scattering cross section, the Stokes parameters,
and -- as as function of the extinction and scattering cross section -- the Albedo.

The weight $w_j(a)$, which represents the contribution of the $j$th component
of the dust grain mixture, results from the abundance of this component
in respect of its dust grain number density (assuming $n_{\rm D}$ chemically
different dust species) and the size distribution of the respective material:
\begin{equation}\label{weight}
  \sum^{n_{\rm D}}_{j=1}\int^{a_{\rm max}}_{a_{\rm min}} w_j(a)\, da = 1\ ,
\end{equation}
where $a_{\rm min}$ and $a_{\rm max}$ are the minimum and maximum grain radius
of the size distribution.
The Stokes parameters as well as the extinction, absorption, and scattering
cross section ($C_{\rm ext}$, $C_{\rm abs}$, $C_{\rm sca}$) are additive. 
Therefore, the representative values in case
of a dust grain mixture can be derived on the basis of their weighted contributions
(see, e.g.,  Martin~1978, \u{S}olc~1980):
\begin{equation}\label{cextmean}
  \langle C_{\rm ext} \rangle = \sum^{n_{\rm D}}_{j=1} \int^{a_{\rm max}}_{a_{\rm min}} 
  w_j(a) \, C_{{\rm ext}_j}(a)\, da,
\end{equation}
\begin{equation}
  \langle C_{\rm abs} \rangle = \sum^{n_{\rm D}}_{j=1} \int^{a_{\rm max}}_{a_{\rm min}} 
  w_j(a) \, C_{{\rm abs}_j}(a)\, da,
\end{equation}
\begin{equation}
  \langle C_{\rm sca} \rangle = \sum^{n_{\rm D}}_{j=1} \int^{a_{\rm max}}_{a_{\rm min}} 
  w_j(a) \, C_{{\rm sca}_j}(a)\, da,
\end{equation}
and
\begin{equation}
  \langle \hat{S} \rangle = \sum^{n_{\rm D}}_{j=1} \int^{a_{\rm max}}_{a_{\rm min}} 
  w_j(a) \, \hat{S}_j(a)\, da,
\end{equation}
where $\hat{S}$ is the Mueller matrix which is used to describe the modification
of the photon's Stokes vector due to the interaction of a photon with the 
scattering/absorbing medium (dust grains; see Bickel \& Bailey~1985, Bohren \& Huffman~1983).
For the albedo $A$ follows:
\begin{equation}\label{albmean}
  \langle A \rangle = \frac{\sum^{n_{\rm D}}_{j=1}\int^{a_{\rm max}}_{a_{\rm min}}
    w_j(a) \, C_{{\rm ext}_j}(a) \, A_j(a)\, da}
  {\sum^{n_{\rm D}}_{j=1}\int^{a_{\rm max}}_{a_{\rm min}}
    w_j(a) \, C_{{\rm ext}_j}(a)\, da}
  =\frac{\langle C_{\rm sca} \rangle}{\langle C_{\rm ext} \rangle}
\end{equation}
Since this formalism (Eq.~\ref{weight}-\ref{albmean})
covers both grain size and chemical distributions, the conclusions to be derived for a 
grain size distribution but a single chemical component is also valid for grain
ensembles consisting of grains with different chemical compositions.
The mean dust grain parameters used in the following sections have been derived
on the assumption of $10^3$ grain radii equidistantly distributed in the grain
size range of the small/large grain ensemble.

In thermal equilibrium, the temperature of a dust grain ($u$th grain size or composition)
can be estimated from the assumption of local energy conservation. If 
\begin{equation}
  dE_u^{\rm abs} = dt \int_0^{\infty} L_{\lambda,u}^{\rm abs}\ d\lambda
\end{equation}
is the energy being absorbed by the $u$th dust grain component during the time interval 
${\rm d}t$, then the local energy conservation can be written as 
\begin{equation}
  dE_u^{\rm ree} 
  = dt \int_0^{\infty} L_{\lambda,u}^{\rm ree}\ d\lambda
  = dt \int_0^{\infty} 4 \pi a^2 \, Q^{\rm abs}_{\lambda,u}(a)
  \, \pi B_{\rm \lambda}(T_u) d\lambda
  = dE_u^{\rm abs},
\end{equation}
where $dE_u^{\rm ree}$ is the total amount of energy being re-emitted during
the time interval $dt$,
$L_{\lambda,u}^{\rm abs}$ and $L_{\lambda,u}^{\rm ree}$
are the monochromatic absorbed/re-emitted luminosity at the wavelength $\lambda$,
$Q^{\rm abs}_{\lambda,u}(a)$ is the absorption efficiency and
$B_{\rm \lambda}(T_u)$ is the Planck function of the grain with the temperature $T_u$.
For two particles of the same composition but different radii ($a_1$, $a_2$)
the temperature is consequently only the same if
\begin{equation}
  Q_{\rm \lambda}^{\rm abs}(a_1) = Q_{\rm \lambda}^{\rm abs}(a_2).
\end{equation}
This is usually not the case.
Thus, mean dust grain parameters may be used only in case of the simulation of
dust scattering but not for the estimation of the dust grain temperature distribution
and resulting observables, such as the spectral energy distribution and
images in the mid-infrared to millimeter wavelength range.

In the following Sections (\S\ref{rtem} and \S\ref{sed}) it will be investigated
under which conditions the approximation of the mean dust grain parameters provides
nevertheless
a good estimate for these observables. While the most crucial information can be 
derived on the basis of a one-dimensional model (\S\ref{rtem}), the study presented
in \S\ref{sed} is aimed to reveal the influence of geometrical effects in case 
of two-dimensional disk-like structures. In each case,
typical parameters for both the dust grain composition (see \S\ref{rtmodel}) 
and the circumstellar shells/disks are considered.

%-----------------------------------------------------------------------------------------------------------
\section{Temperature Distribution}\label{rtem}

\subsection{Basic Model: Spherical Shell}\label{spsh}

In order to reveal the differences between the results of the RT based on real dust grain size
distributions and the approximation of mean dust grain parameters representing the weighted
mean optical parameters of the grain size distribution, a simple-structured
but in respect of the astrophysical applications very useful model has been used for the following
investigations. It consists of a spherical shell (outer radius: $10^3$\,AU;
inner boundary: $0.2$\,AU) with a radial density profile described by a power law,
$\rho(r) \propto r^{\alpha}$,
where $\alpha$ is a negative, constant quantity. The dust grain size distribution is assumed
to be constant throughout the whole shell. The only heating source is an embedded, isotropically
radiating star in the center of the shell with an effective temperature and radius identical to that
of the Sun.

In order to study the influence of multiple scattering and (additional) heating
due to the dust reemission, the optical depth of the shell $\tau$ (as seen from the star)
and the exponent $\alpha$, describing the relative density gradient in the shell,
have been considered as variable parameters of the model. 
Four values (-1.0, -1.5, -2.0, and -2.5) for the exponent $\alpha$
were chosen, covering a broad range of astrophysical objects ranging from 
circumstellar shells (see, e.g., Chini et al.~1986, Henning~1985, Yorke~1980)
and disks (see, e.g., Men'shchikov \& Henning~1997, Sonnhalter et al.~1995, Lopez et al.~1995) 
to the radial density profile being assumed for the dust density
distribution in AGN models (see, e.g., Manske et al.~1998,
Efstathiou \& Rowan-Robinson~1995, Stenholm~1994). 
The optical depth $\tau$ was varied from the optical
thin case ($\tau$=0.1) to the case of intermediate optical depth ($\tau=1$) and finally to
$\tau=10$ as the optically thick case. For optical depths below this interval no remarkable 
differences to the case $\tau=0.1$ are expected since multiple scattering and reemission have
a negligible influence on the dust grain temperature which is determined only by the distance
from the star (attenuation of the stellar radiation field) and the absorption efficiency
of the dust grains. For optical depths $\tau \gg 10$ preparatory studies have shown that
the temperature difference between the different dust grains becomes negligible, converging
towards the temperature distribution obtained on the basis of the mean dust grain approximation.
This finding is in agreement with results obtained by Kr\"ugel \& Walmsley~(1984 -- 
investigation of dust and gas temperatures in dense molecular cores),
Wolfire \& Churchwell~(1987 -- study of circumstellar shells around low-mass, pre-main-sequence
stars), and
Efstathiou \& Rowan-Robinson~(1994 -- multi-grain dust cloud models of compact H\,II regions).
Furthermore, only the inner region of the circumstellar shell shows a broad distribution 
of dust grain temperatures at a given distance from the star. 
\begin{figure}[t]
  \epsscale{1.0}
  \plotone{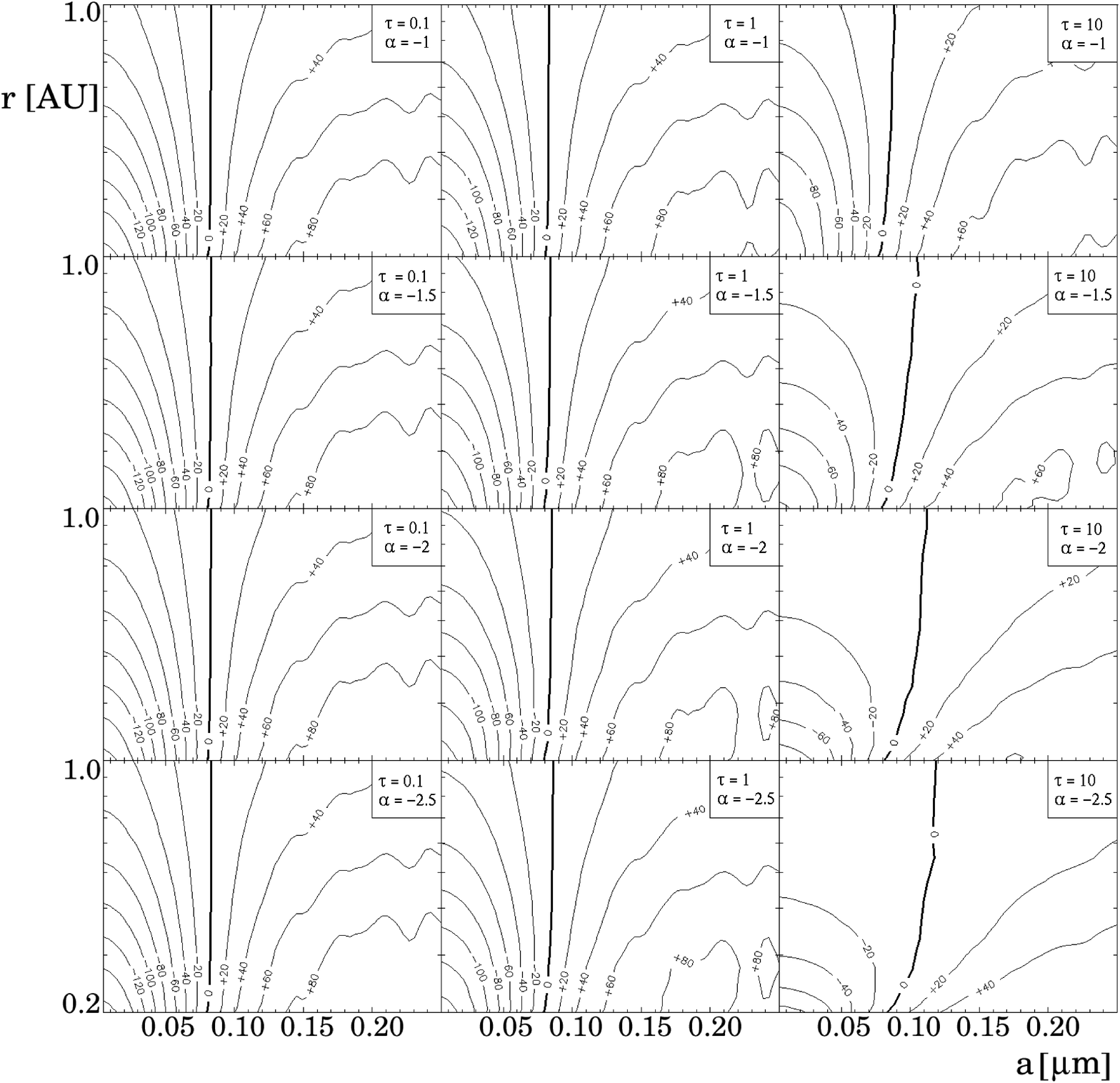}
  \caption{Temperature difference of the dust grains $T(a,r)-\bar{T}(r)$ 
    in the innermost region of the shell (inner
    boundary at 0.2\,AU--1\,AU). The quantity $T(a,r)$ is the radial temperature
    distribution of the dust grains with the radius $a$; 
    $\bar{T}(r)$ is the radial temperature distribution resulting from the approximation
    of the mean dust grain parameters. Small dust grain size distribution: $a=(0.005-0.25)\,\mu$m.
    The number of grain sizes being considered amounts to~$2^6$.
  }
  \label{s6txax2}
\end{figure}
\begin{figure}[t]
  \epsscale{1.0}
  \plotone{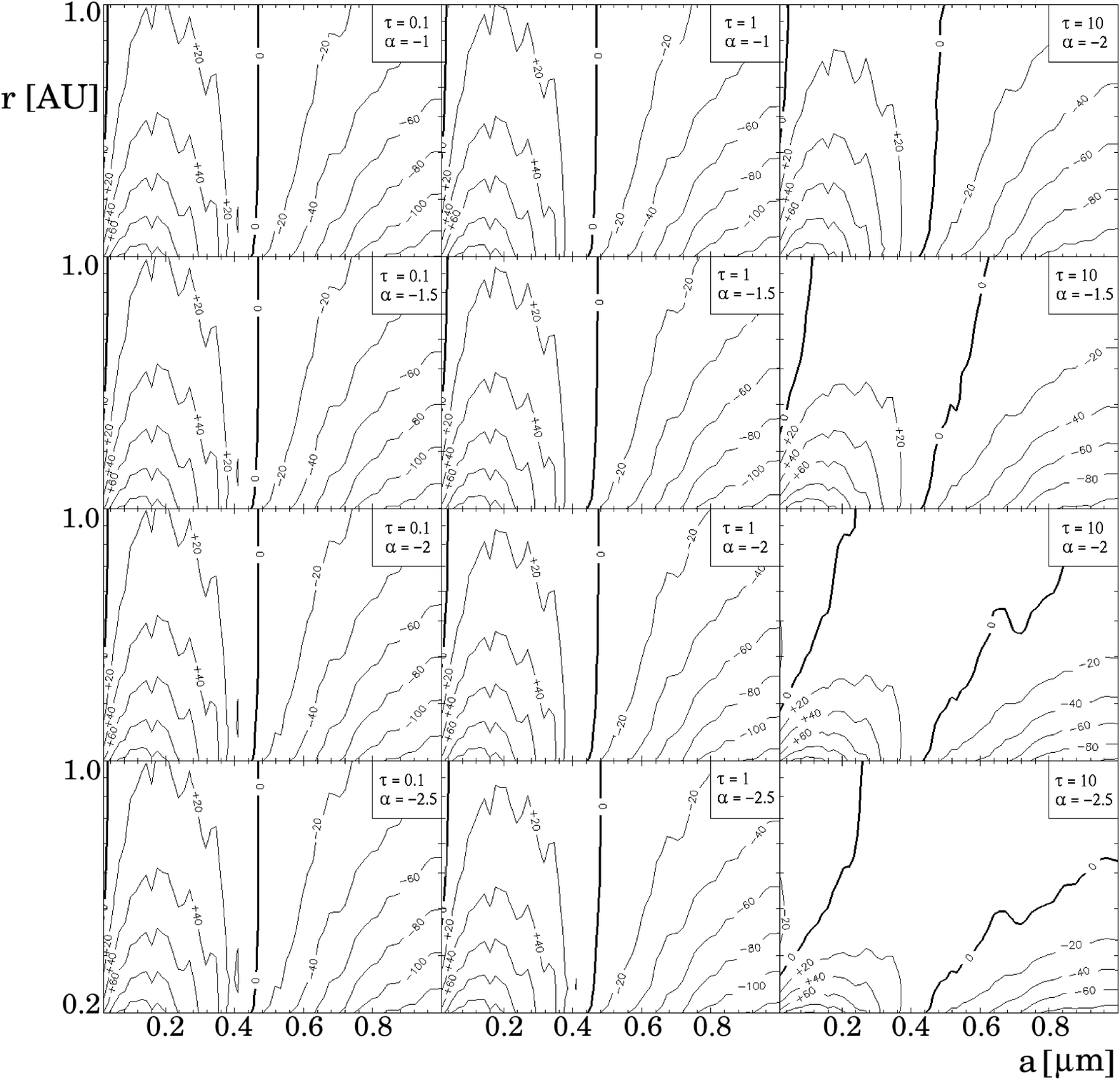}
  \caption{Temperature difference of the dust grains $T(a,r)-\bar{T}(r)$ 
    in the innermost region of the shell (inner
    boundary at 0.2\,AU--1\,AU).
    Large dust grain size distribution: $a=(0.1-1)\,\mu$m. 
    Note that the temperature of the dust grains
    mainly increases with decreasing particle radius, which is in contrast to the 
    behaviour of the small grain size distribution (see Fig.~\ref{s6txax2}; 
    see Sect.~\ref{spsh} for explanation).
  }
  \label{s6ta-b2}
\end{figure}
In Fig.~\ref{s6txax2} and \ref{s6ta-b2} the differences between the radial temperature distribution
of the single dust grains $T(r,a)$ and that resulting from the approximation of mean dust grain
parameters $\bar{T}(r)$ is shown
from the inner boundary of the shell to a distance of 1\,AU from the star.
The main results are:
\begin{enumerate}
\item The temperature of the different grains spans a range of $\approx250$\,K ($\approx25$\%
of the corresponding mean temperature) around the temperature of the mean grains 
at the inner boundary of the shell.
\item The temperature difference decreases
	\begin{itemize}
	\item towards larger distances from the star, and
	\item with increasing optical depth as soon as the shell becomes optically thick.
	\end{itemize}
\item The relativ density gradient, described by the exponent $\alpha$ is of minor importance.
However, an increase of the absolute amount of this exponent results in an increased redistribution of energy
between the different dust grain components at the inner boundary of the shell ($\tau \ge 1$)
and therefore in a decrease of the temperature dispersion.
\item The difference ($T(r,a) - \bar{T}(r)$) strongly depends on both the absorption efficiency,
and thus on the dust grain size distribution, and the temperature
of the heating source since the combination of these parameters determines the amount of energy
being absorbed. This is clearly illustrated by the different
signs of the temperature difference of the large and the small grain size distribution
to the temperature of the mean grains shown in Fig.~\ref{s6txax2} and \ref{s6ta-b2}.
\end{enumerate}
The found complex, highly dispersed temperature structure in a dust grain mixture
requires a large number (here: $2^6$) of considered grain radii
to be correctly taken into account
and cannot be adequately represented by a mean particle. This finding is
of special interest for the simulation of chemical networks, 
 since the existence of ice layers,
 the possibility of a certain reaction on the grain surface,
 surface reaction rates, 
 and the temperature of the surrounding gas phase
depend on the dust grain temperature (see, e.g., 
Cazaux \& Tielens~2002,
Markwick et al.~2002,
Charnley~2001).
The dust phase in the close stellar environment is of particular importance,
since the chemical evolution takes place here on its smallest timescale.
Furthermore it was found, that the approximation of mean particle properties is found to be justified
for dust configurations of high ($\tau \gg 10$) optical depth.

%...........................................................................................................
\subsection{HH\,30 circumstellar disk}\label{hh30}

Although all basic characteristics of the temperature structure in a dust grain mixture can be studied
on the basis of a one-dimensional model (Sect.~\ref{spsh}), a disk-like structure will be considered
in the following. The motivation for this is to investigate the possible influence of geometrical
effects which have been minimized in case of the spherical shell.
Furthermore, two-dimensional models are of particular importance since they are widely applied
in simulations of circumstellar disks, 
debris disks around evolved stars, 
active galactic nuclei, and
galaxies (see, e.g., Chiang \& Goldreich~1997, 1999;
see also the references in Sect.~\ref{spsh}).

In the following, a model of the nearly edge-on disk around the classical T-Tauri star HH\,30
is considered (see, e.g., Burrows et al.~1996 for observational details).
Following Cotera et al.~(2001) and Wood et a.~(2002), a flared geometry as described by
Shakura \& Sunyaev~(1973) is adapted:
\begin{equation}
  \rho = \rho_0 
  \left(
  \frac{R_*}{{\tilde \omega}}
  \right)^\alpha
  \exp
  \left(
  -\frac{1}{2}
  \left[
    \frac{z}{h({\tilde \omega})}
    \right]^2
  \right),
\end{equation}
where ${\tilde \omega}$  is the radial coordinate in the disk midplane, $\rho_0$ is the density
at the stellar surface, and the scale height $h$ increases with radius:
\begin{equation}
  h = h_0 
  \left(
  \frac{\tilde \omega}{R_*}
  \right)^\beta.
\end{equation}
In order to reproduce the high density (and therefore temperature) gradient in the inner
region of the disk, a very high resolution of the model both in radial and vertical direction
has been applied. The smallest resolved structure at the inner radius of the disk has 
a linear extent of about 29\,\% both in vertical and radial direction.
%\begin{figure}[t]
%  \epsscale{1.0}
%  \plotone{f3.eps}
%  \caption{Radial resolution of the HH\,30 model. The model space is subdivided in volume elements
%    bothin radial and vertical direction which represent the smallest size scale on which 
%    the temperature gradient can be simulated 
%    (see Wolf, Henning, \& Stecklum~1999 for a detailed description).
%    The smallest resolved structure at the inner radius of the disk has a linear extent of about 29\,\%
%    of the stellar radius.
%    The rectangular boxes mark the half number of cells (in radial direction) within the
%    areas of the disk shown in Fig.~\ref{disk-t} (corresponding to the radius of the mapped regions
%    of 0.2, 2, 20, and 200\,AU). The subdivision of the model space in $\theta$ direction
%    has been adapted to the radial subdivision in order to achieve a similar resolution both
%    in radial and vertical direction at each point in the disk.}
%  \label{grid}
%\end{figure}
As for the SED modelling performed by Wood et al.~(2002), the following values have been used:
stellar radius $R_* = 1.2 R_\sun$, 
stellar effective temperature $T_* = 3500 {\rm K}$,
$\alpha = 2.25$,
$\beta = 1.25$,
$h_0 = 0.017 R_*$,
inner radius of the disk $R_0 = 6 R_* $, and 
the outer radius of the disk amounts to 200\,AU.
In contrast to the simulations presented by Cotera et al.~(2001) and Wood et al.~(2002),
only one chemical dust grain component (astronomical silicate as in Sect.~\ref{spsh}) has been chosen
in order to simplify the simulation analysis.
The size distribution is specified using a power law with exponential decay:
\begin{equation}
n(a) \propto a^{-p} 
\exp
\left(
-\frac{a}{a_{\rm c}}
\right)
\end{equation}
(see, e.g., Kim et al.~1994).
Following Cotera et al.~(2001), a maximum grain size $a_{\rm max} = 20 \mu{\rm m}$ has been
chosen (other parameter values: $a_{\rm c}=0.55 \mu{\rm m}$, $p = 3.0$). 
A total number of 32 grain sizes with radii equidistantly distributed on a logarithmic
scale between 0.05\,$\mu{\rm m}$ and 20\,$\mu{\rm m}$ are considered.
The visual extinction of the disk amounts to $A_{\rm V} = 2800$ at an inclination of $84^{\rm o}$
(see Wood et al.~2002, Tabl.~2). The RT simulations have been performed on the basis
of the model geometry shown in Fig.~6[B] in Wolf, Henning, \& Stecklum~1999.
While the star could be assumed to be point-like in the one-dimensional model (\S\ref{spsh}),
its real extend has to be taken into account now. The radiation characteristic at each point
of the stellar surface is described by a $I \propto \cos(\theta)$ law
(star without limb-darkening; see Wolf~2002 for details).

\begin{figure}[t]
  \epsscale{1.0}
  \plotone{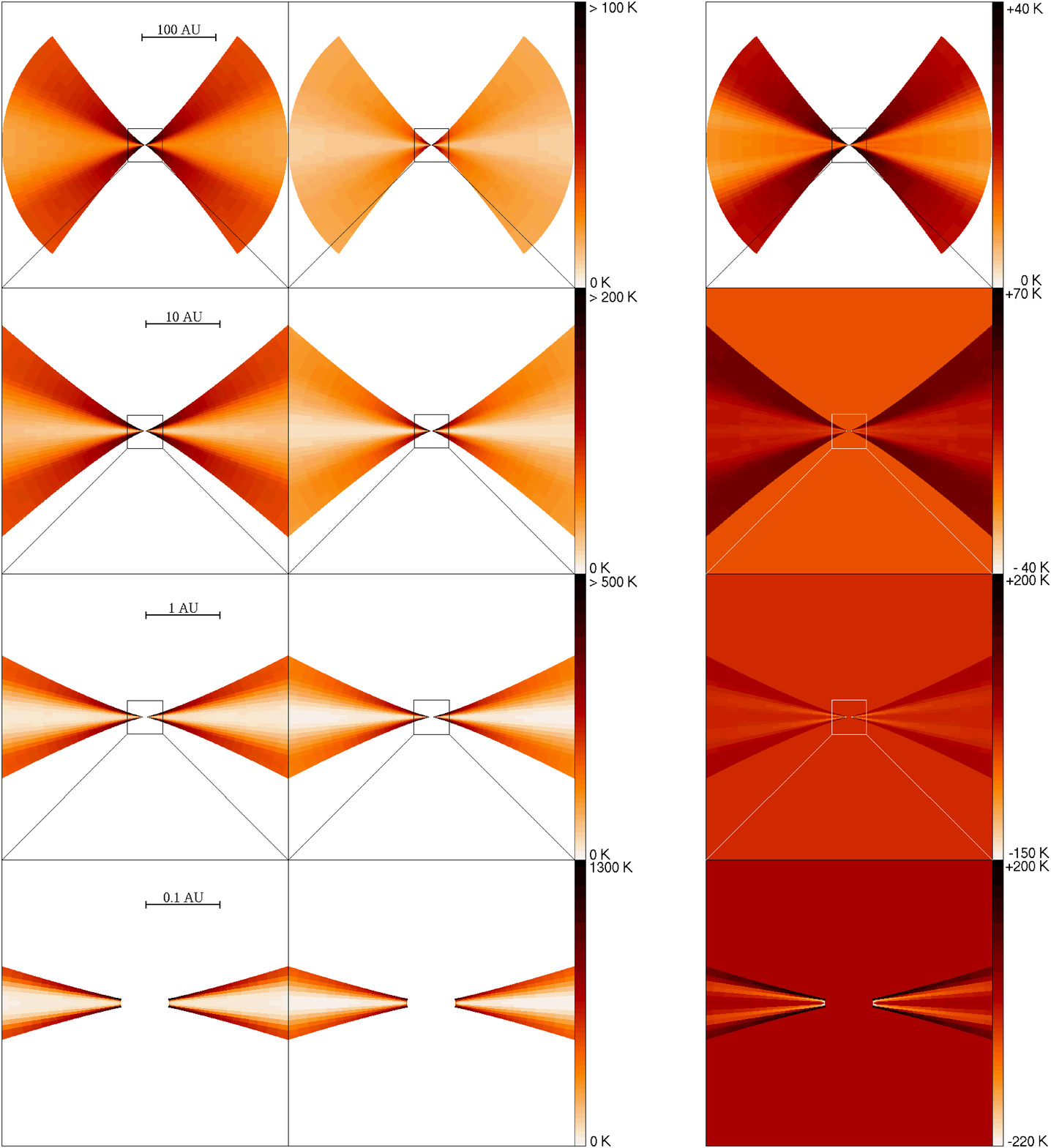}
  \caption{
    {\sl Left and middle column:} 
    Color coded temperature structure of the HH\,30 circumstellar disk model
    (left: $a=0.05 \mu{\rm m}$ dust component,
    middle: $a=20 \mu{\rm m}$ component).
    {\sl Right column:} 
    Temperature difference $\Delta T = T(a=0.05 \mu{\rm m}) - T(a=20 \mu{\rm m})$. 
    The background color in the dust free regions above/below the disk represents the
    $\Delta T = 0$ level. Thus, regions with brighter colors than the background color represent
    areas where the large grains have a higher temperature than the small grains
    (see also Fig.~\ref{disk-z}).
    {\em [See the electronic edition of the Journal for a color version of this figure.]}
  }
  \label{disk-t}
\end{figure}
The resulting temperature distribution for two selected grain sizes
($a=0.05 \mu{\rm m}$ and $a=20 \mu{\rm m}$) and their difference are shown in Fig.~\ref{disk-t}.
On the largest, $\sim$10--100\,AU scale the temperature of the small grains is up to 40\,K
higher than that of the large grains in the optically thin atmosphere above/below the disk.
In agreement with the results found in optically thick spherical shells,
the temperature difference decreases towards the optically extremly thick midplane
where it amounts to only a few Kelvin.

\begin{figure}[t]
  \epsscale{0.38}
  \plotone{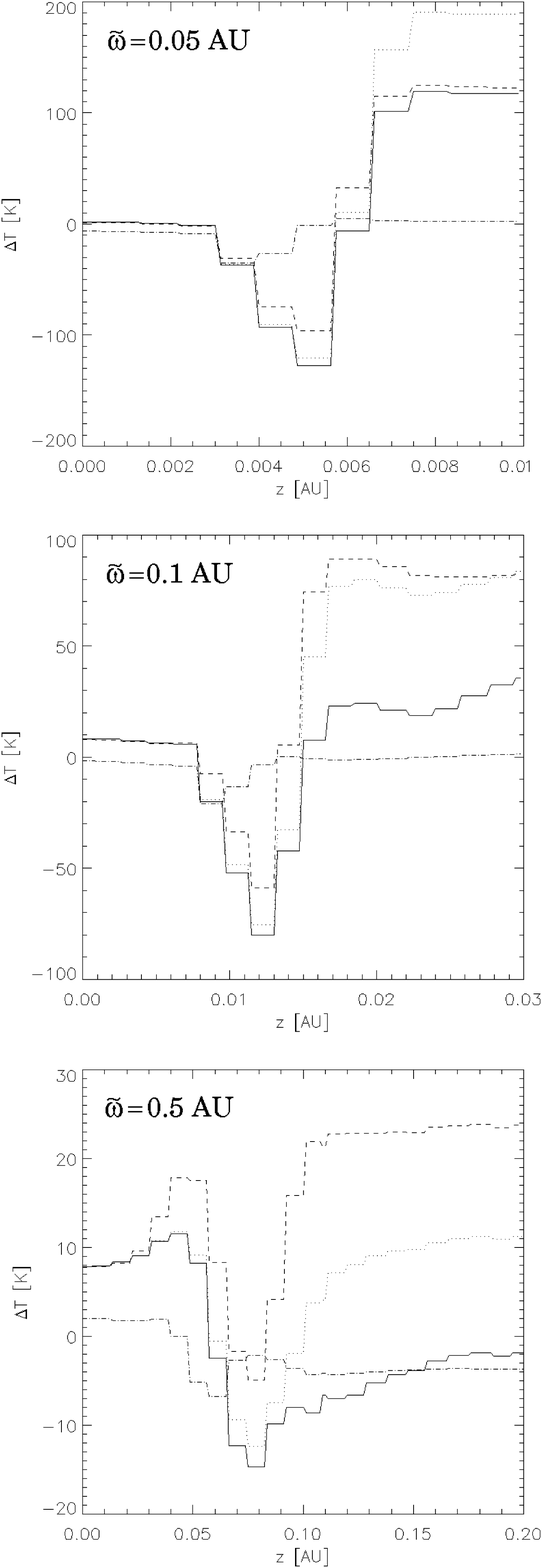}
  \caption{Temperature difference $\Delta T = T(a) - T(a=20\mu{\rm m})$ as a function
    of the distance from the midplane $z$, obtained at radial distances
    $\tilde{\omega}$=0.05\,AU, 0.1\,AU, and 0.5\,AU from the star.
    Considered particle sizes:
    $a$ = 
    0.01\,$\mu{\rm m}$ (solid line),
     0.1\,$\mu{\rm m}$ (dotted line),
     1.1\,$\mu{\rm m}$ (dashed line), and
     9.0\,$\mu{\rm m}$ (dash-dotted line).
  }
  \label{disk-z}
\end{figure}
\begin{figure}[t]
  \epsscale{0.38}
  \plotone{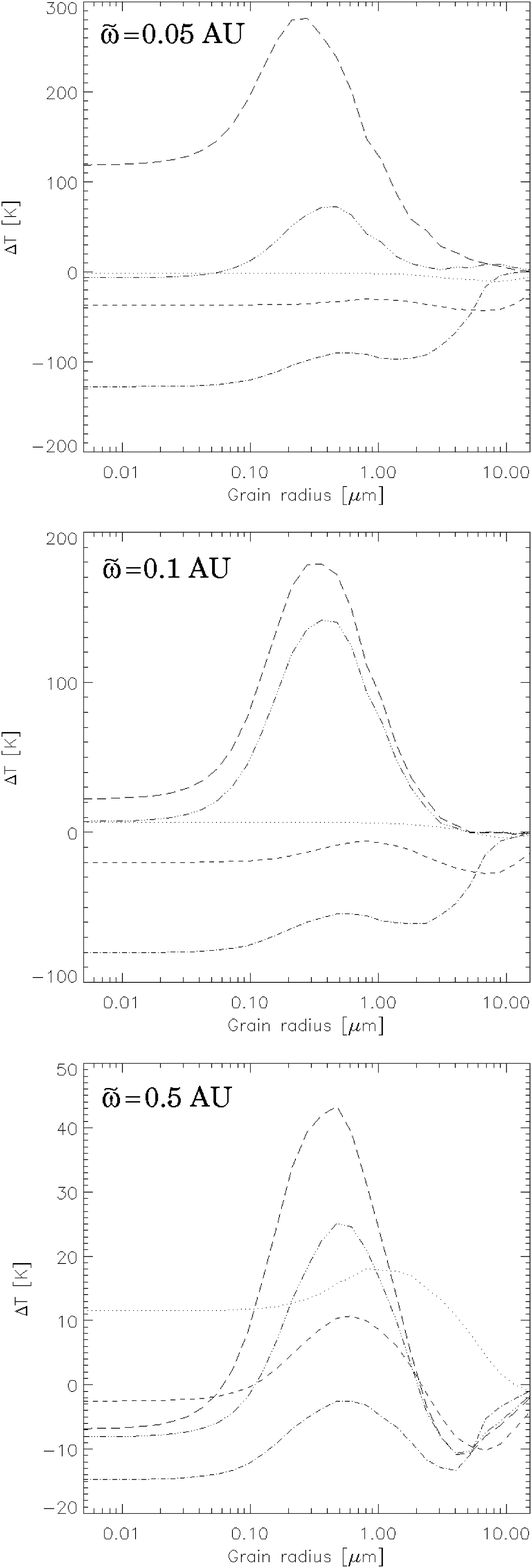}
  \caption{Temperature difference $\Delta T = T(a) - T(a=20\mu{\rm m})$ as a function
    of the grain radius $a$ obtained at radial distances
    $\tilde{\omega}$=0.05\,AU, 0.1\,AU, and 0.5\,AU from the star.
    In each figure, the temperature difference $\Delta T$ is shown at 5 points
    with different distances from the midplane $z$ which are located relative
    to the coordinate $z=z_{\rm m}$ at which the minimum of the temperature difference
    for $a$ = 
    0.01\,$\mu{\rm m}$,
    0.1\,$\mu{\rm m}$ (dotted line), and
    1.1\,$\mu{\rm m}$
    was found in Fig.~\ref{disk-z}:
    $z = 0.5 z_{\rm m}$ (dotted line),
    $z = 0.75 z_{\rm m}$ (short-dashed line),
    $z = z_{\rm m}$ (dot-dashed line),
    $z = 1.25 z_{\rm m}$ (dot-dot-dot-dashed line), and
    $z = 1.5 z_{\rm m}$ (long-dashed line).
  }
  \label{disk-g}
\end{figure}
However, the vertical temperature structure shows a more complex behaviour in the inner
region of the disk, on a size scale of a few AU and smaller. 
Here, an {\em inversion} of the temperature difference in the upper layer of the optically 
thick disk occurs, i.\,e., the large grains are substantially warmer than the small grains.
This remarkable effect can be explained by the more efficient heating of large grains by the
mid-infrared to far-infrared reemission of the hot inner disk.
This explanation is also supported by the comparison of the vertical temperature structure
obtained at different distances from the star\footnote{In order
to obtain the spatially high-resolved temperature distributions shown in Fig.\ref{disk-z}
and \ref{disk-g} on a PC with 1\,GByte RAM, the RT has been simulated only in the inner
disk region with a diameter of 8.8\,AU, which is one to two orders of magnitude larger than
the regions considered in the following discussions. The comparison with the results obtained
on a temperature grid with a lower resolution (Fig.~\ref{disk-t}) showed that the resulting
differences in the spatial temperature distribution is in the order of or smaller
than the statistical noise of the results and therefore negligible.}: 
Fig.~\ref{disk-z} shows the 
difference between vertical temperature distribution of 
grains with increasing radii 
($a=0.01\,\mu{\rm m}$, 0.1\,$\mu{\rm m}$, 1.1\,$\mu{\rm m}$, 9.0\,$\mu{\rm m}$)
and $a=20\,\mu{\rm m}$ grains. Furthermore, Fig.~\ref{disk-g} shows
the temperature difference $\Delta T = T(a) - T(a=20\mu{\rm m})$ as a function
of the grain radius $a$ and radial distance from the star at different distances
from the midplane.
While at a radial distance of 0.05\,AU the temperature of the 20\,$\mu$m grains is about 130\,K
higher than that of $0.01\,\mu{\rm m}$ grains in the ``inversion layer'',
the amount of this temperature difference drops to less than 
$\sim 20 {\rm K}$ at a radial distance of 0.5\,AU from the star (see Fig.~\ref{disk-z}).
Consequently, the vertical extent and location of the minimum of the temperature inversion region
depends on the particular grain size (and the vertical density distribution in the disk).
For instance it was found that the minimum of the $\Delta T = T(a=9.0\mu{\rm m}) - T(a=20\mu{\rm m})$
distribution shown in Fig.~\ref{disk-z} is shifted towards smaller distances from the midplane
in all three considered cases ($\tilde{\omega}$=0.05\,AU, 0.1\,AU, and 0.5\,AU).
According to the explanation of the temperature inversion given above,
a similar shift of the minima of the difference temperature distribution is expected
in case of the other grain sizes (0.01\,$\mu{\rm m}$ to 1.1\,$\mu{\rm m}$) as well.
However, the spatial distribution of the simulated temperature distribution is too low 
to allow the verification of this assumption.

%In the optically thin atmosphere, the temperature structure returns to ``normal''
%conditions, i.e., the smaller grains are hotter than the larger grains.

%-----------------------------------------------------------------------------------------------------------
\section{Spectral Energy Distribution}\label{sed}

\begin{figure}[t]
  \epsscale{1.0}
  \plotone{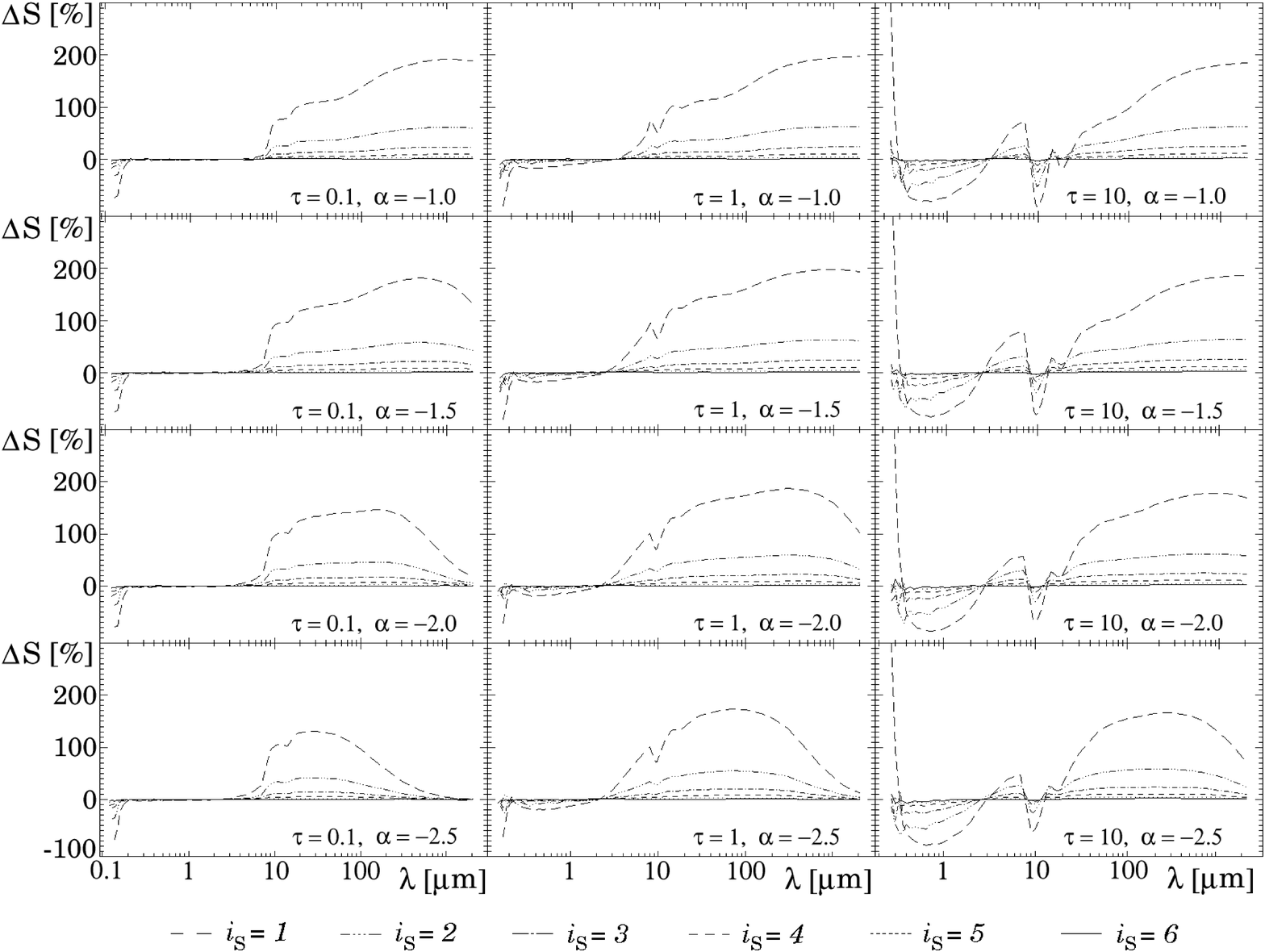}
  \caption{Relative difference $\Delta S_\lambda$ of the SEDs resulting from a real dust grain mixture 
    based on $2^{i_{\rm S}}$ ($i_{\rm S} = 1, 2,\ldots, 6$) dust grain sizes and SEDs
    based on the approximation of mean dust grain parameters.
    See the upper left Figure for the explanation of the different
    drawing style. A close-up view on the difference spectrum $\Delta S_\lambda$ in the case of
    $2^6$ grain sizes is shown in Fig.~\ref{sed2}.}
  \label{sed1}
\end{figure}

\begin{figure}[t]
  \epsscale{0.5}
  \plotone{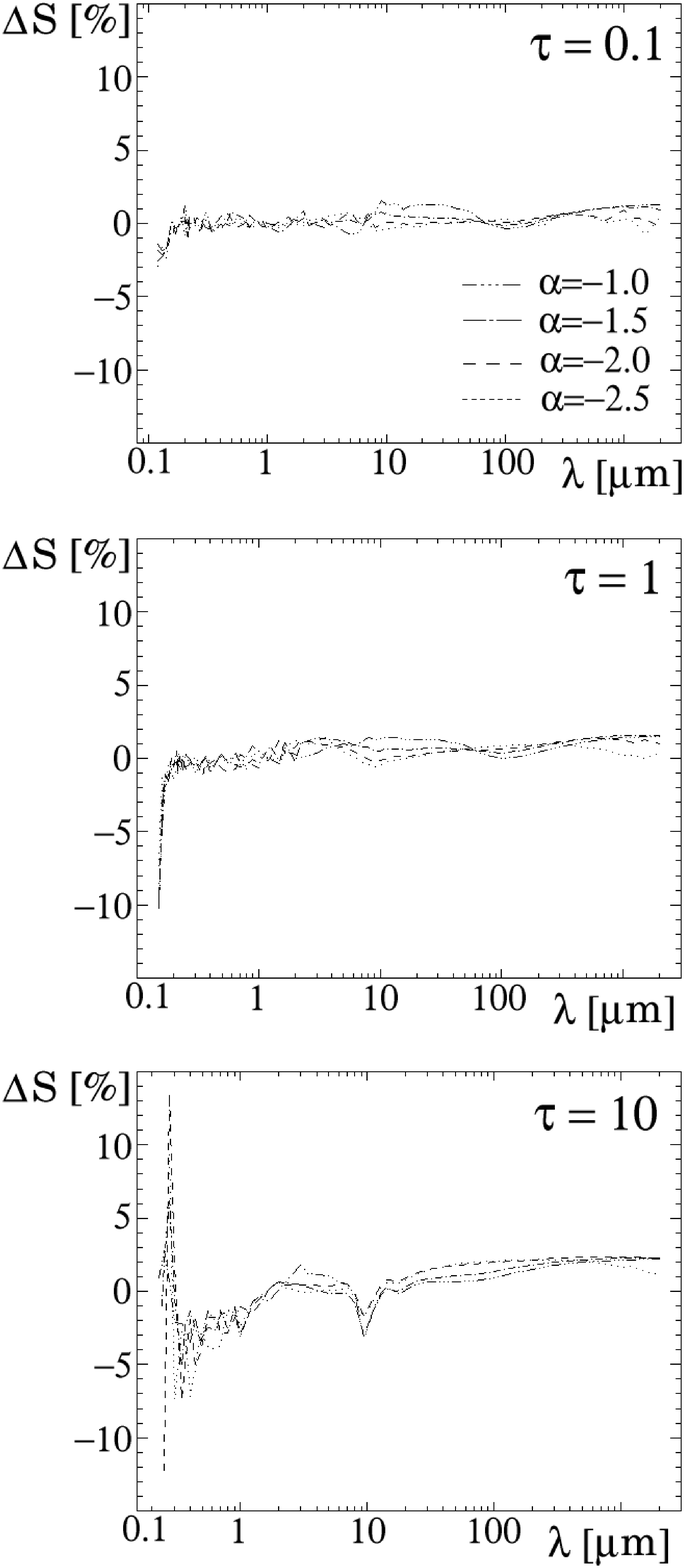}
  \caption{Relative differences $\Delta S_\lambda$ between the SEDs based on $2^6$ dust grain sizes
    and the SED based on the approximation of mean dust grain parameters.
    The influence of the exponent $\alpha$
    (see \S\ref{spsh}) is clearly of minor importance.}
  \label{sed2}
\end{figure}

Based on the dust grain temperature distribution shown in Fig.~\ref{s6txax2},
the spectral energy distribution (SED) for the different (spherical) shells
have been calculated assuming $n_{\rm S} = 2^{i_{\rm S}}$ ($i_{\rm S} = 1, 2,\ldots, 6$)
different grain sizes. The relative difference
\begin{equation}
  \Delta S_\lambda = 
  \frac{S_\lambda(n_{\rm S}) - \bar{S}_\lambda}{\bar{S}_\lambda}
\end{equation}
is shown in Fig.~\ref{sed1} and \ref{sed2}.
Here, $S_\lambda(n_{\rm S})$ represents the SED of the real grain size distribution,
and $\bar{S}_\lambda$ is the SED resulting from the approximation of mean dust grain parameters.

As Fig.~\ref{sed1} illustrates, the results of the real grain size distribution
converges towards the results based on the approximation of mean dust grain parameters.
As it also shows, a minimum of $2^5$ to $2^6$ grain sizes have to be considered 
to achieve deviations of less then $\sim 10\%$. If too few single grain sizes are considered, 
the SED does not represent the observable SED
of the real grain size distribution sufficiently but drastically overestimates
\begin{enumerate}
\item The absorption of the stellar radiation
  (and therefore underestimates the visual to near-infrared SED),
\item The depths of the absorption bands in general (as the silicate absorption
  feature at $\sim 9.7\,\mu$m demonstrates), and
\item The dust reemission spectrum in the near-infrared to millimeter
  wavelength range.
\end{enumerate}

While the different absorption behaviour influences the temperature structure and overall
SED in general, the influence of the different scattering behaviour depends on the optical depth.
In the optically thin case, differences in the scattering behaviour result mainly in deviations
of the short wavelength region of the stellar SED. At higher optical depth, the scattering
behaviour has direct influence on the energy transfer in the dust envelope and thus on the temperature
structure and the near- to far-infrared wavelength range as well.
However, this effect is of importance only at intermediate optical depths ($\tau\approx1\ldots10$),
since for even higher optical depths the temperature structure, at least in case of a one-dimensional
dust configuration, tends not to depend on the description of the dust grain ensemble 
(real dust grain size distribution / approximation of mean dust grain parameters).

%-----------------------------------------------------------------------------------------------------------
\section{Conclusions}\label{concl}

In simulations of the radiative transfer in the circumstellar environment of young stellar objects 
it has been widely established to use mean values for those parameters 
which describe the interaction of the electromagnetic field with the dust grains
(see, e.g., modelling efforts by
Calvet et al.~2002: circumstellar disk around the young low-mass star TW Hya;
Cotera et al.~2001 and Wood et al.~2002: circumstellar disk around the classical T~Tauri star HH~30;
Fischer, Henning, \& Yorke~1994, 1996: polarization maps of young stellar objects).
A main reason for this is given by the fact that the consideration of a
large number of single grain sizes and chemically different grain species
results in a nearly linearly increasing amount of required computer memory in order
to store the separate temperature distributions. Furthermore, the calculation of
the spatial temperature distribution for all grains of different sizes and chemical
composition requires significantly more computing power since 
(a) the heating by the primary sources (e.g., the star embedded in a circumstellar envelope)
has to be performed independently for each grain species, and
(b) the number of computing steps required to model the subsequent mutual heating 
of the different dust grain species due to dust re-emission scales
even as $\approx (n_{\rm D} \times n_{\rm S})^2$, where 
$n_{\rm D}$ is the number of chemically different components and
$n_{\rm S}$ is the number of separate dust grain radii considered.
While these simulations are feasible in case of one-dimensional models
(see, e.g.,
Chini, Kr\"ugel, \& Kreysa~1986 and Efstathiou \& Rowan-Robinson~1994: 
Dust emission from star forming region;
Kr\"ugel \& Siebenmorgen~1994 and Siebenmorgen, Kr\"ugel, \& Zota~1999: Radiative transfer in galactic nulei)
or simple-structured two-dimensional models 
(Men'shchikov \& Henning~1997: Circumstellar disks;
Efstathiou \& Rowan-Robinson~1994: Disks in active galactic nuclei),
it is hardly possible to handle two- and three-dimensional models 
with high density gradients and/or high optical depth
which require high-resolution temperature grids.

In this study, the difference between the results of RT simulations based 
(a) on a mean dust grain parameter approximation and 
(b) real dust grain size distributions 
have been investigated. 
Based on a one-dimensional density distribution it was found that the temperature
structure of a real grain size distribution shows a very complex behaviour
in the inner, hot region of the shell depending on
(1) the grain size distribution,
(2) the effective temperature of the embedded star and the optical depth
and therefore on the density distribution. However, the relative difference between
the SED based on a real dust grain size distribution on the one hand and the approximation
of mean dust grain parameters on the other hand was found to be smaller than about 
$\approx$10\% if a minimum number of $2^5$ to $2^6$ grain sizes have been considered.

As the temperature structure in a circumstellar disk --
based on the model for HH\,30 -- shows, the geometry of the density distribution
is a significant parameter for the resulting temperature differences between grains
of different size, too. In the inner region of the disk with a diameter
of a few AU a temperature inversion layer was found where the sign of the temperature
difference of the largest and smallest grains is reverted.
As this and the results obtained on the basis of the spherical shell (\S\ref{spsh})
show, the dust grain temperature structure is not sufficiently represented by dust
grains with mean optical parameters.
On the one hand, this is of tremendous importance for the simulation of chemical networks
since the largest deviations from results based on the approximation of mean dust grain 
parameters have been found in the inner hot, dense region of the shell/disk 
where the chemical evolution takes place on its smallest
timescale. On the other hand, the complex temperature structure may significantly
change the hydrostatic properties of the considered gas/dust density distribution
itself. However, these questions have to be investigated in future studies in order
to find out the influence on observable quantities such as SEDs, images, polarization
maps, and visibilities.
Furthermore, the influence on processes taking place in more evolved 
circumstellar disks, such as the dust settling and dust grain growth, have to be
considered taking into account the temperature structure of real grain size distributions.

%-----------------------------------------------------------------------------------------------------------
\acknowledgments
This research was supported through the HST Grant GO\,9160, 
and through the NASA grant NAG5-11645.
I wish to thank the referee E.\ Dwek who helped to improve the clarity
of the presentation of the results.

%% Appendix material should be preceded with a single \appendix command.
%% There should be a \section command for each appendix. Mark appendix
%% subsections with the same markup you use in the main body of the paper.

%% Each Appendix (indicated with \section) will be lettered A, B, C, etc.
%% The equation counter will reset when it encounters the \appendix
%% command and will number appendix equations (A1), (A2), etc.

%\appendix

%\section{Dust grain parameters}

\end{document}